\documentclass{aa}
\usepackage{natbib}
\usepackage{aas_macros}
\usepackage{graphicx}

\bibpunct{(}{)}{;}{a}{}{,}

\def\4u{4U 1626-67}
\def\apgt{\ {\raise-.5ex\hbox{$\buildrel>\over\sim$}}\ }
\def\aplt{\ {\raise-.5ex\hbox{$\buildrel<\over\sim$}}\ }                     %
\newcommand{\pyr}{\mbox {{\rm yr$^{-1}$}}}
\newcommand{\rs}{\mbox {$R_{\odot}$}}
\newcommand{\ms}{\mbox {$M_{\odot}$}}
\newcommand{\myr}{\mbox {$M_{\odot}\,{\rm yr^{-1}}$}}
\newcommand{\md}{\mbox {$\dot{M}$}}

\begin{document}
\title{On the formation of neon-enriched donor stars in ultracompact X-ray binaries}

\author{L. R. Yungelson\inst{1,2}
        \and 
        G. Nelemans\inst{3}
         \and 
           E. P. J. van den Heuvel\inst{2}
       }
\offprints{L. Yungelson, lry@inasan.rssi.ru}

\institute{Institute of Astronomy of the Russian Academy of
             Sciences, 48 Pyatnitskaya Str., 119017 Moscow, Russia  
         \and
           Astronomical Institute ``Anton Pannekoek'', 
             Kruislaan 403, NL-1098 SJ Amsterdam, the Netherlands 
             \and
              Institute of Astronomy, University of Cambridge,
              Madingley Road, CB3 0HA, Cambridge, UK 
          }

\date{Received  21 December 2001 /Accepted 3 April 2002}

\titlerunning{Formation of neon-enriched donors in X-ray binaries}
\authorrunning{Yungelson, Nelemans \& van den Heuvel }

\abstract{We study the formation of neon-enriched donor stars in ultracompact X-ray
  binaries $(P_{\mathrm{orb}} < 80$\,min) and show that their progenitors have to be  low-mass (0.3 -- 0.4)\,\ms\  ``hybrid'' white dwarfs (with CO
  cores and thick helium mantles). Stable mass transfer is possible if in the initial
  stages of mass exchange mass is lost from the system, taking away
  the specific orbital angular momentum of the accretor (``isotropic
  re-emission''). The excess of neon in the transferred matter is due
  to chemical fractionation of the white dwarf  
which has to occur prior to the Roche lobe overflow by the donor. The estimated  lower limit of the orbital periods of the systems with neon-enriched donors is close to 10 min. 
  We show that the X-ray pulsar 4U 1626-67, which likely also has a
  neon-enriched companion, may have been formed via accretion induced
  collapse of an oxygen-neon white dwarf accretor if the donor was a
  hybrid white dwarf.
}

\maketitle

\section{Introduction}
\label{sec:intro}
Recently, \citet{jpc00} reported the discovery of the excess of neon
in four low-mass X-ray binaries: 4U 0614+091, 2S 0918-549, 4U
1543-624, and 4U 1850-0857. It is suggested that the Ne is local to
these objects. 4U 1850-0857, which belongs to a globular cluster, has
a measured orbital period of 20.6 min. From the optical properties of
the systems, Juett et al.  suggest that all of them are ultracompact:
$P_{\mathrm{orb}} < 80$\,min. A local excess of neon is found also in
the ultracompact ($P_{\mathrm{orb}} = 41.4$\,min) binary pulsar 4U
1626-67 \citep{ang+95,sch+01}.  Juett et al. propose that in all these
binaries the donors are low-mass, neon-rich degenerate dwarfs.
\citet{sch+01} suggest that the donor in 4U 1626-67 is the core of a
carbon-oxygen (CO) or oxygen-neon (ONe) white dwarf which has
previously crystallised. However, the evolution of neutron star+white dwarf [henceforth, (wd,ns)] binaries  to the Ne-enrichment stage was not studied. 

Below, we explore the evolution of white dwarfs which fill their Roche lobes having neutron star companions.  We discuss the
mechanism of mass exchange and limiting masses of the donors in these binaries in
Sect. \ref{sec:evol}. The process of chemical fractionation that could
lead to Ne-enrichment is briefly outlined in Sect. \ref{ssec:neon}.
The observed systems are discussed in Sect. \ref{sec:obs}.
Some uncertainties of the model and related problems are considered in Sect. \ref{sec:disc}. Our conclusions follow in Sect. \ref{sec:concl}.

\section{Evolution of ultracompact binaries} 
\label{sec:evol}
\subsection{Formation  of ultracompact binaries}
\label{ssec:form}

After  pioneering work of \citet{pac67} on the ultrashort-period
($\sim\!\!18$\,min) cataclysmic variable AM CVn, it is commonly accepted that
the donors in semi-detached systems with the orbital periods of several tens of
minutes may be degenerate dwarfs and that the driving force of the binary
evolution is angular momentum loss (AML) due to gravitational wave radiation
(GWR).

The basic features of the scenario for the formation of   (wd,ns) systems
which can evolve into ultracompact X-ray binaries may be summarized as follows \citep[see e. g.][ for details]{ty93c,ity95b}.
In the Galactic disk 
stars with  masses $ >\!10$\,\ms\ that are born in binaries evolve
into neutron stars, experiencing underway one or two common envelope
stages which strongly decrease the orbital separation.  Next, their companions experience dynamically unstable mass
loss and become white dwarfs.  If
the resulting systems have orbital
separation $\sim\!\rs$, AML via GWR may bring the white dwarf to Roche-lobe overflow (RLOF)
within the lifetime of the Galactic
disk and under certain conditions (see below) stable mass
exchange leading to exposure of the Ne-enriched core may ensue.

In globular clusters neutron stars most probably acquire
their companions via exchange interactions with primordial binaries
\citep[e. g.][]{rpr00}. The subsequent evolution is similar to that
of the disk binaries.

\subsection{Mass transfer in ultracompact binaries: the nature of the donors and the limits of their initial mass} 
\label{ssec:exch}
\begin{figure}[t!]
\resizebox{\columnwidth}{!}{\includegraphics[angle=-90]{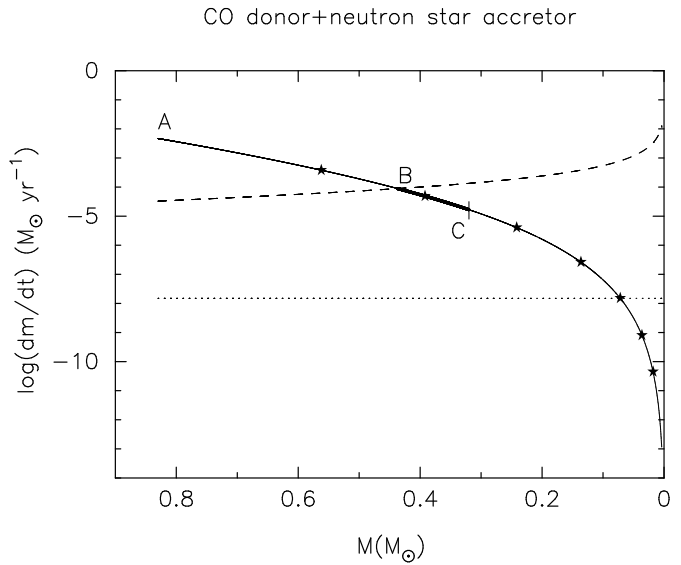}}
\caption[]{The solid line shows  the  mass loss rate from a white dwarf
``stabilised'' by isotropic re-emission  for the  case of a system which
initially contained a 0.83\,\ms\ white dwarf donor and a 1.433\,\ms\ neutron
star accretor.  Dashed line: the upper limit of the mass loss rate from the
system by the isotropic re-emission (for $r_{\mathrm{ns}} = 10$\,km).  Dotted line:
$\md_{\mathrm{Edd}}$\ for a neutron star. Asterisks mark the parameters of the
system at $\log t{\mathrm{(yr)}}$ =3, 4, 5, 6, 7, 8, 9.  Systems with different
initial $M_{\mathrm d}$ follow curves which are indistinguishable on the scale
of the figure. To the left of the beginning of the mass-loss curve (A) the
Roche-lobe filling white dwarfs are dynamically unstable. Between A and B
isotropic re-emission is not efficient enough for ejection of all matter lost
by white dwarf.  ``Hybrid'' white dwarfs which are able to transfer matter
stably have initial masses corresponding to the thick part of the mass-loss
line BC.}
\label{fig:rate}
\end{figure}

In systems with degenerate donors, at the onset of
Roche-lobe overflow the mass transfer rate \md\ may be several orders of
magnitude higher than the Eddington limiting
accretion rate ($\md_{\mathrm{Edd}}$) for a white dwarf or neutron star accretor \citep{vil71,ty79a}, [see Fig.
\ref{fig:rate}]. Formation of a
common envelope engulfing the whole system and merger of components or
collapse of the neutron star into a black hole may be avoided if  
the excess of the
matter which cannot be accreted leaves the system taking away the
specific orbital angular momentum of the accreting component
[``isotropic re-emission'' \citep{sph97}]. This role may be accomplished  by radiatively driven outflows \citep{kb99,ts99}. The equation for the mass
loss rate by a white dwarf in the presence of AML by GWR and ``isotropic re-emission'' used in this study may be found in \citet{ylttf96}.

Figure \ref{fig:rate} shows this mass loss rate
for a (wd,ns) system with initial masses of donor and accretor
$M_{\mathrm d}=0.83\,\ms$ and $M_{\mathrm a}=1.433\,\ms$,
respectively\footnote{We neglect  the effects of an X-ray irradiation induced
stellar wind which may enhance mass loss by the donor \citep{tl93}.}.  Since
all white dwarfs approach  zero-temperature  radii after cooling for several
100 Myr \citep{pab00} and since most systems will be very much older than this
when their orbits have decayed sufficiently to start mass transfer, we use in
our computations the mass -- radius  $(M-R)$  relation  for cold spheres
derived by \citet{zs69} in the form given  by \citet{rap+87}, assuming  equal
mass  fractions of carbon and oxygen. For this $M-R$  relation, if the mass
retention efficiency of the accretor is close to zero, (wd,ns)  systems with 
$M_{\mathrm d} \apgt 0.83$\,\ms\ are dynamically unstable.

All matter in excess of $\md_{\mathrm{Edd}}$\ can be lost from the system only
if the liberated accretion energy of the matter falling from the Roche lobe
radius of the neutron star to the neutron star surface is sufficient to expel
the matter from the Roche-lobe surface around the neutron star, i.e. $\md <
\md_{\mathrm{max}} = \md_{\mathrm{Edd}}\,(r_{\mathrm{L,ns}}/r_{\mathrm{ns}})$, where
$r_{\mathrm{ns}}$\ is the radius of the neutron star \citep{kb99,ths00}.  The
dependence  of $\md_{\mathrm{max}}$\ on the mass of the donor   is also plotted
in Fig.  \ref{fig:rate}. Since the mass-loss line in Fig.~\ref{fig:rate}  holds
as well for initial $M_{\mathrm d} < 0.83\,\ms$, within our assumptions
``isotropic re-emission'' from a neutron star is actually possible for  donors
with initial  $M_{\mathrm d} \aplt 0.44\,\ms$. Unless there is an even more
efficient way to stabilise the mass transfer than by ``re-emission'', this
excludes ONe white dwarfs as donors as these have $M \apgt 1.1\,\ms$\
\citep{gpgb01}. Similarly also massive CO dwarfs are excluded as donors.

The overabundance (relative to solar) of Ne 
suggests, that the progenitors of the donors in the X-ray systems under
consideration have experienced core helium burning including the
$\mathrm{^{14}N(\alpha,\gamma)^{18}F(\alpha,\gamma)^{22}Ne}$ reactions
chain. Thus the donors cannot be low-mass helium white dwarfs.

This leaves as the last option the so called ``hybrid''  white dwarfs which have CO cores and thick He mantles.
They are formed from components of close binaries with
initial masses in the range
from 2.5 to 5\,\ms\ which experience RLOF prior to core He ignition, 
become He-burning stars and, after completion of core He burning, evolve directly into white dwarfs. 
Their mass is $\simeq(0.3 - 0.7)$\,\ms\ \citep{it85,hte00}.

In a calculation of the model of the population of compact
stars in the Galaxy \citep{nyp01} the birthrate of such (wd,ns)
systems with $M_{\mathrm{wd}} \leq  0.4$\,\ms\ is $2.7\times
10^{-6}$\,\pyr. For comparison, the birthrate of progenitors of
X-ray binaries with hydrogen-rich donors in the same model is $1.3\times
10^{-5}$\,\pyr.

\subsection{Neon-enrichment}
\label{ssec:neon}

\begin{figure}[ht!]
\resizebox{\columnwidth}{!}{\includegraphics[angle=-90]{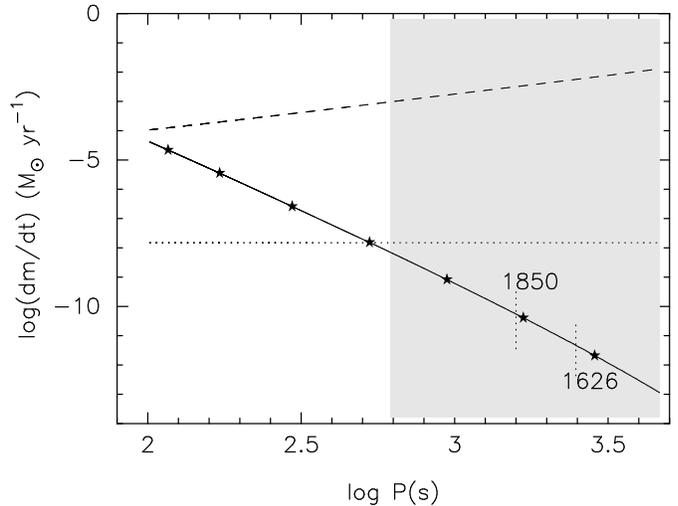}}
\caption[]{Orbital period -- mass loss rate relation for a system which
  initially contained a 0.38\,\ms\ ``hybrid'' white dwarf donor
  and a 1.433\,\ms\ neutron star accretor. The meaning of the
  lines and symbols is as in Fig.\,\ref{fig:rate}. The
  vertical dashed lines mark the orbital periods of 4U 1850-0857
  and 4U 1626-67. Within the period range corresponding to the
  shaded area the donors may transfer neon-enriched matter.}
\label{fig:pmdot}
\end{figure}

Even after core helium exhaustion the abundance of Ne in the cores of hybrid
white  dwarfs is low, so to observe the strong Ne enrichment, the cores have to
be   crystallised and fractionated. These  processes take  several Gyr,
depending on the  mass of the dwarf, transparency of the outer layers, total
amount of ${\mathrm{^{22}Ne}}$ and uncertainties in phase diagrams
\citep{her+94,mkww99}.  Figure \ref{fig:pmdot} shows that  the orbital periods
of (wd,ns) pairs  enter the observed range of $\sim 10$\,min in less than
10\,Myr. (The meaning of the lines and asterisks in Fig. \ref{fig:pmdot} is the
same as in Fig. \ref{fig:rate}).  Hence, the enrichment of the white dwarf core
by Ne, must  have happened before RLOF.

For an initial metallicity Z=0.02, after
completion of He-burning, 
the mass abundance of $\mathrm{^{22}Ne}$
in the core of the star cannot be larger than $X_{\mathrm{^{22}Ne}} = 0.02$. 
During crystallisation of \textit{binary} mixtures of  (C/Ne) and (O/Ne), Ne settles in the center  and forms a Ne-enriched nucleus which 
contains all $\mathrm{^{22}Ne}$\ formed in the dwarf and has the so called azeotropic abundance $X_{\mathrm{{^{22}}Ne}}^{\mathrm a}$ of Ne \citep{ise+91}.

The mass of this Ne-rich nucleus is
\begin{equation}  
\label{eq:necore}
M_{\mathrm{Ne-r}}=\left(X_{\mathrm{^{22}Ne}}/X_{\mathrm{^{22}Ne}}^{\mathrm
a}\right) M_{\mathrm{wd}}. 
\end{equation}
If, following \citet{bicl92}, one assumes
$X_{\mathrm{^{22}Ne}}^{\mathrm a}=0.07$ --  an intermediate value of
$X_{\mathrm{^{22}Ne}}^{\mathrm a}$ for O/Ne and C/Ne mixtures -- and takes for
$M_{\mathrm{wd}}$\ the maximum mass of the convective core of a 0.378\,\ms\
``hybrid'' dwarf -- 0.2\,\ms\ \citep{it85}, then the mass $M_{\mathrm{Ne-r}}$\
of the Ne-enriched core is $\simeq 0.06$\,\ms. 

If the white dwarf didn't have enough time to crystallise, the mass loss
uncovers material of the convective core of the progenitor of the dwarf.   Then
the  mass fraction of Ne may be down  to the initial $\sim 0.001$.

\section{Comparison with the observed Ne-rich X-ray binaries}
\label{sec:obs}
\subsection{4U 0614+091, 2S 0918-549, 4U 1543-624, and 
  4U 1850-0857}

Figure \ref{fig:pmdot} shows the evolution of a semidetached system
which initially contained a ``hybrid'' white dwarf and a neutron star,
in orbital period -- mass loss rate coordinates. Taking the mass of the Ne-enriched core of 0.06\,\ms\ estimated in Sect.
\ref{ssec:neon} as typical, one obtains a rough lower limit of
the orbital periods of binaries with the Ne-rich white dwarf donors of
about 10\,min. The white dwarf enters the range of periods (or masses)
for which Ne-enrichment is possible in several Myr
following the onset of RLOF
(Fig.~\ref{fig:pmdot}).

Our calculations of the mass loss rate also give the mass of the donor
as a function of the orbital period.  We thus infer the donor masses for 4U
1850-0857 and 4U 1626-67 to be 0.027 and 0.01\,\ms, respectively (see more detailed discussion of the latter system in Sect. \ref{ssec:4u}).

The abundance of Ne relative to solar found in the observed Ne-rich X-ray
binaries ranges from $1.9\pm0.3$\ to $2.87\pm0.16$. For oxygen an
underabundance is reported: $0.37\pm0.06$\ to $0.52\pm0.04$. Taken at face
value, these relative abundances translate into a local Ne/O ratio of 0.63 --
0.77.  If true, these values are  considerably higher than one would expect for
the  above used  abundance of Ne in the nucleus of the dwarf 
$X_{\mathrm{^{22}Ne}}^{\mathrm a} = 0.07$. But \citet{ise+91} notice, that
$X_{\mathrm{^{22}Ne}}^{\mathrm a}$\  may be well underestimated by a factor
$\sim 3$.  If this is true, the Ne/O ratio comes into better agreement with
observed values. The mass of the Ne-enriched core then becomes  slightly higher
than 0.02\,\ms, still comparable  with our  estimates of the masses of the
white dwarfs in 4U 1850-0857 and 4U 1626-67.

The problem of a low predicted Ne/O ratio would also hold for initially more massive donors,
if stable mass loss would be possible for them. In the core of a
1.1\,\ms\ ONe dwarf the Ne/O ratio is $\sim0.4$\ \citep{gpgb01}.
For a ``standard'' 0.6\,\ms\ CO white dwarf with an initially
equimolar distribution of C/O and traces of Ne, a final Ne/O ratio
$\sim 0.25$\ is expected in the core \citep{sch+94}.

All four Ne-enriched X-ray binaries contain, presumably, weakly
magnetised neutron stars. This may be a result of the decay of the
magnetic field by accretion \citep{th86}.

\subsection{4U 1626-67: an accretion induced collapse?}
\label{ssec:4u}
4U 1626-67 differs from the binaries discussed  above by the presence of a
7.7\,s X-ray pulsar with a strong magnetic field estimated from cyclotron
emission: $(3.2\pm 0.1)(1+z)\!\times\!10^{12}$\,G, where $z$\ is gravitational
redshift \citep{orl+98}.  Applying a polytropic $M-R$ relation, \citet{vwb90}
estimated its donor mass as $\sim 0.02\,\ms$. A similar estimate  of $M_{\mathrm
d}$ was obtained by \citet{sch+01} under the {\it assumption} that velocities 
of the lines in the Ne/O complex and their Doppler shifts reflect Keplerian
motions in the accretion disk. Thus, existing estimates of $M_{\mathrm d}$ are
consistent with Ne-enrichment of the donors. A very low value of $M_{\mathrm d}$\
in 4U 1626-67 is also favoured by the probability of detection considerations
\citep{vwb90}.

The high magnetic field strength suggests that the neutron star is young and
accretion was negligible.  The absence of enhanced abundances of O-group
elements in the spectrum of 4U 1626-67 argues against the origin of the neutron
star in a SN\,II event \citep{ang+95}.  \citet{th86} suggested that this X-ray
pulsar was formed recently -- $(2 - 10)\times10^7$\, yrs ago -- by an accretion
induced collapse (AIC) of a white dwarf. \citet{vwb90} rejected the AIC model,
based on the argument that  the time between AIC and resumption of mass
transfer is $\sim 10^8$\,yr, longer than then assumed e-folding decay time
($\sim 10^7$\,yr) for the magnetic fields of neutron stars. Later, the 
analysis of magnetic fields of isolated radio pulsars have shown that magnetic
field decay scale is probably $\apgt 10^8$\,yr \citep{bwhv92}. This, along with
the discovery of Ne-enrichment in the system, suggests that one should 
reconsider the formation of 4U 1626-67 via AIC, especially, since the details
of pre-collapse evolution were never studied before. 

In the model for the population of compact stars in the Galaxy \citep{nyp01}
the  birthrate of ONe white dwarfs with $M\geq 1.2$\,\ms\ accompanied by
``hybrid'' white dwarfs with $ 0.35 \leq M \leq 0.5$\,\ms\ which get into
contact within 10 Gyr is $1.25\times 10^{-6}$\,\pyr. This rate is within the
limits for occurrence of AIC's in the Galaxy set by nucleosynthesis
considerations \citep{fbhc99}.

\begin{figure}[t!]
\resizebox{\columnwidth}{!}{\includegraphics[angle=-90]{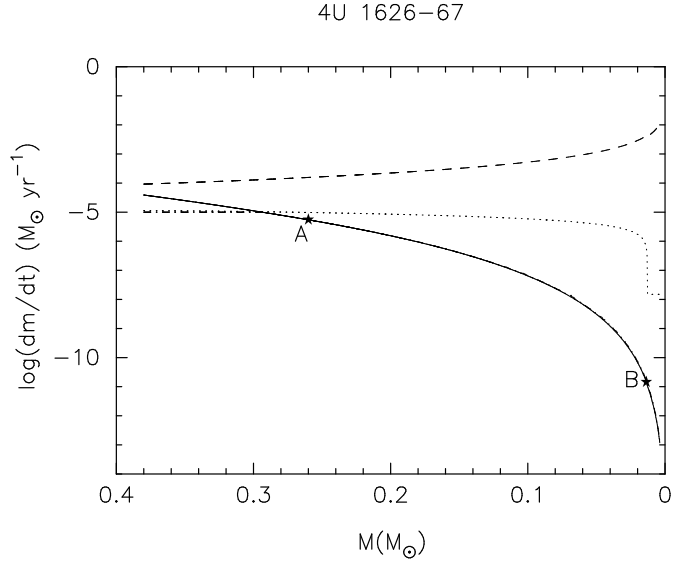}}
\caption[]{Solid line -- mass exchange rate in a system  which
  initially consisted of a 0.38\,\ms\ ``hybrid'' donor and a 1.2\,\ms\ 
  ONe white dwarf accretor. The dashed line is the upper limit of
  the ``isotropic re-emission'' rate for an accreting white
  dwarf. The dot-dashed line -- accretion rate. The dashed line --
  $\md_{\mathrm{Edd}}$, first for a white dwarf, then for a neutron star.
  At point A the He mantle of the donor is exhausted, at point B
  the white dwarf collapses into a neutron star. }
\label{fig:4u}
\end{figure}

The initial binary may contain e. g. a 0.38\,\ms\  ``hybrid'' white dwarf [for
which we know the internal structure from \citet{it85}] and a 1.2\,\ms\
oxygen-neon one (Fig.\,\ref{fig:4u}). At the beginning of the RLOF 
$P_{\mathrm{orb}}=2$\,min. The outer 0.12\,\ms\ of the donor consist of He.  The maximum
rate of stable He-burning at the surface of a massive white dwarf  is about
$10^{-5}\,\myr$ \citep{kh99}, lower than the mass loss rate by the donor in the
initial stages of mass transfer.     Mass exchange  in this case may be also
stabilised by isotropical re-emission of optically thick wind from accretor,
generated by He-burning \citep{hkn96}.  Therefore stable helium burning is
likely to set in on the surface of the 1.2\,\ms\ accreting white dwarf.

Like \citet{vwb90},  we assume that the white dwarf collapses into a 1.26\,\ms\
neutron star after the increase in mass to 1.44\,\ms. This happens after
240\,Myrs of accretion, when $P_{\mathrm{orb}}\! \approx\!2000$\,s. The mass of
the donor is at that moment 0.0132\,\ms.  The Ne/O ratio in the core for
azeotropic abundance of Ne (0.1 to 0.3, see \ref{ssec:neon}) doesn't contradict
the measured Ne/O ratio in 4U 1626-67 \citep{sch+01}: $0.22\pm0.15$.  

The collapse interrupts mass transfer for $\sim 1.56\times10^8$\,yr due to the
loss of binding energy of the dwarf \citep[see for details][]{vwb90}. This time
span may not be sufficient for the decay of the magnetic field of the pulsar. 
When after resumption of the contact the period of the system has increased to
41.4\,min, the mass of the donor is 0.0101\ms.

Until this moment the neutron star in our model system has
lived after the AIC for about 350 Myr.  If magnetic fields
do not decay on long time scale its field will not have  decayed sufficiently.
The accretion of 0.003\,\ms\ onto a neutron star is most probably
insufficient for destruction of its magnetic field (if  the frozen
field and incompressible fluid approximations \citep{cz98} are assumed). Assuming solid body rotation and
conservation of angular momentum, one gets $\simeq 1.3$\,ms\ for the
initial spin period of the neutron star. The present spin
period of the neutron star in 4U 1626-67 is 7.7 s, close to the
``death-line'' of radio pulsars for $B=3\times 10^{12}$\,G\ 
\citep{bha96}. Thus, it is conceivable that 4U 1626-67 harbours a
neutron star which ceased to be a radio pulsar, but didn't  spin
down sufficiently to experience recycling.

At the stage when our model system resembles 4U 1626-67,
the mass exchange rate is $\simeq\!4\times10^{-12}\,\myr$. As already noticed before \citep{chak98,sch+01}, all model \md\ for  4U 1626-67 based on $M-R$ relations are
considerably lower than
the estimates of \md\ based on the observed spin-up rate and simple
angular momentum conservation considerations.
It is plausible that  the secular \md\ is consistent
with model expectations, while mass transfer rate \md\ inferred from the observed
spin-up of the pulsar (if correct) reflects the accretion rate from an
unstable accretion disk \citep{sch+01}. The latter    experiences thermal ionization instability if 
$\md_{\mathrm a} \! \aplt \! 7.4 \! \times \! 10^{-10}$ and
$\aplt \! 2.5 \! \times \! 10^{-9}$\,\myr, for a pure C or O disk, respectively \citep{mph02}.

\section{Discussion}
\label{sec:disc}
White dwarfs  may be 
considered as a \textit{ternary} C/O/Ne ionic mixture. For such mixtures  
\citet{og+93} predict the formation of an almost pure Ne core upon
solidification for \textit{any} C:O ratio. If this core ultimately
contains all $\mathrm{^{22}Ne}$\ generated in the white dwarf its mass
is only $\sim 0.004\,\ms$, much less than given by Eq.\,(\ref{eq:necore}) and this would require much larger $P_{\mathrm{orb}}$\ for X-ray binaries showing Ne enrichment.

On the other hand, according to
\citet{segr96}, if a 0.6\,\ms\ white dwarf is considered as  a ternary mixture with $X_{\mathrm C}=X_{\mathrm O}=0.495,
X_{\mathrm{^{22}Ne}}=0.01$, almost all Ne concentrates in a thin layer around $M_r/M_{\mathrm{wd}} \approx 0.7$. Then spilling of Ne-enriched
matter over the neutron star would require extremely fine tuning of
the model.

Another intriguing problem concerns X-ray bursts. They were reported
for 4U 0614+091, 2S 0918-549, and 4U 1850-0857. 
None of the bursts have shown atypical behaviour \citep{jpc00}.
Carbon flashes at $\md < 0.1 \md_{\mathrm{Edd}}$\ were never studied.
Extrapolation of the calculations of \citet{cb01} over 2 orders of
magnitude may be not very relevant. 
However, it shows
that for $X_{\mathrm C} \sim 0.2$, characteristic for 
``hybrid'' dwarfs and $\md \sim 10^{-11}$\,\myr, accreted carbon
possibly  burns stably. Thus, the
origin of the bursts in systems with  \textit{any} low-mass C-rich donors   
deserves further attention.

\section{Conclusion}
\label{sec:concl}

We have shown that within the framework of the white dwarfs cooling
model developed by
 \citet{ise+91}, \citet{bicl92}, \citet{sch+94}, \citet{her+94}
 it is possible to explain the formation of the
Ne-enriched donors in  X-ray binaries if their progenitors were low-mass CO white
dwarfs with  thick He mantles. A necessary condition is that the
white dwarfs had enough time to be substantially chemically
fractionated. This means that the mass transfer had to start several
Gyr after the formation of the white dwarfs. We estimate that the masses of Ne-rich donors in these systems are $\aplt 0.06\,\ms$ and their orbital periods have to be $\apgt 10$\,min. 
The expected Ne/O ratios are $\sim(0.1 - 0.4)$.

Finally we conclude that the X-ray pulsar 4U 1626-67 could follow an evolutionary path similar to the other Ne-rich systems, with the only
difference that it used to have a massive ONe white dwarf component that collapsed into a neutron star relatively recently.
We estimate the current mass of the donor in this system as 0.01\,\ms.     

\begin{acknowledgements}
We acknowledge fruitful discussions with A. Kudrjashoff, O. Pols, G.-J.
Savonije, E. Garc\'{i}a-Berro, J. Isern, R. Ramachandran.  Our thanks a    due
to Z. Han for communicating his unpublished results.  LRY acknowledges the warm
hospitality and support of the Astronomical Institute ``Anton Pannekoek''. This
work was supported by a NOVA grant, NWO Spinoza grant 08-0 to E. P. J. van den
Heuvel,   and ``Astronomy'' Program. 
\end{acknowledgements}

\bibliography{h3381}
\end{document}